\RequirePackage{silence}
\ErrorFilter{babel}{You haven't loaded the option}
\ErrorFilter{babel}{You haven't defined}
\pdfoutput=1

\documentclass[aip,pof,reprint,twocolumn,floatfix]{revtex4-1}

\usepackage{graphicx}
\usepackage{amsmath}
\usepackage[table]{xcolor}
\usepackage{colortbl}
\usepackage{color}
\definecolor{darkblue}{rgb}{0,0,0.5}
\definecolor{darkred}{rgb}{0.5,0,0}
\definecolor{darkgreen}{rgb}{0,0.3,0}

\usepackage[colorlinks=true,linkcolor=darkblue,anchorcolor=black,citecolor=darkred,
menucolor=black,urlcolor=darkblue,runcolor=black,filecolor=darkgreen,
plainpages=false,pdfpagelabels=true,breaklinks,letterpaper=true]{hyperref}

\begin{document}


\title{Bi-stability in turbulent, rotating
spherical Couette flow
}


\author{Daniel S. Zimmerman}
\email[email address: ]{danzimmerman@gmail.com}
\author{Santiago Andr\'{e}s Triana}
\email[email address: ]{triana@umd.edu}
\affiliation{Department of Physics, \\ Institute for Research in
Electronics and Applied Physics, University of Maryland, College
Park, MD 20742}
\author{D. P. Lathrop}
\email[email address: ]{lathrop@umd.edu}

\affiliation{Department of Physics, Department of Geology\\ Institute for Research in
Electronics and Applied Physics, University of Maryland, College
Park, MD 20742}


\date{\today}

\begin{abstract}
Flow between concentric spheres of radius ratio $\eta = r_\mathrm{i}/r_\mathrm{o} = 0.35$ is studied in a 3~m outer diameter experiment. We have measured the torques required to maintain constant boundary speeds as well as localized wall shear stress, velocity, and pressure. At low Ekman number  $E = 2.1\times10^{-7}$ and modest Rossby number $0.07 < Ro < 3.4$, the resulting flow is highly turbulent, with a Reynolds number ($Re=Ro/E$) exceeding fifteen million.  Several turbulent flow regimes are evident as $Ro$ is varied for fixed $E$. We focus our attention on one flow transition in particular, between $Ro = 1.8$ and $Ro = 2.6$, where the flow shows bistable behavior. For $Ro$ within this range, the flow undergoes intermittent transitions between the states observed alone at adjacent $Ro$ outside the switching range.   The two states are clearly distinguished in all measured flow quantities, including a striking reduction in torque demanded from the inner sphere by the state lying at higher $Ro$.  The reduced angular momentum transport appears to be associated with the development of a fast zonal circulation near the experiment core.  The lower torque state exhibits waves, one of which is similar to an inertial mode known for a full sphere, and another which appears to be a strongly advected Rossby-type wave.   These results represent a new laboratory example of the overlapping existence of distinct flow states in high Reynolds number flow. Turbulent multiple stability and the resilience of transport barriers associated with zonal flows are important topics in geophysical and astrophysical contexts.

\end{abstract}


\maketitle


\section{\label{intro}Introduction}

\par An understanding of rapidly rotating turbulent flow is key to
many problems of geophysical and astrophysical fluid dynamics. The large
scale fluid motions of stars as well as planetary atmospheres, oceans, and cores
are strongly influenced by the Coriolis
accelerations arising from the rapid overall rotation in those systems.
These flows can be described by the Navier-Stokes equation written
for a frame rotating with constant angular velocity
$\mathbf{\Omega} = \Omega\hat{z}$, which in dimensionless form is

\begin{equation}
\label{rotNS} \frac{\partial\mathbf{u}}{\partial{t}} + Ro\
\mathbf{u}\cdot\nabla\mathbf{u} + 2\hat{z}\times\mathbf{u} =
-\nabla p + E\ \nabla^2\mathbf{u}.
\end{equation}
Using $U$ as the characteristic velocity scale, $\ell$ as the
characteristic length scale, and the kinematic viscosity $\nu$, we
define the relevant dimensionless parameters.  The Rossby number,
\begin{equation}Ro = \frac{U}{\Omega \ell},\end{equation} expresses
the strength of the nonlinear term relative to the Coriolis term
$2\hat{z}\times\mathbf{u}$, and the Ekman number,
\begin{equation}E = \frac{\nu}{(\Omega \ell^2)},\end{equation}
characterizes the importance of viscous drag relative to Coriolis
acceleration.  Throughout this paper we use, $\ell =
r_\mathrm{o}-r_\mathrm{i}$, $\Omega =
\Omega_\mathrm{o}$, and $U =
(\Omega_\mathrm{i} - \Omega_\mathrm{o})\ell$, so that
\begin{equation}
\label{Rodef}   Ro = \frac{\Delta\Omega}{\Omega_\mathrm{o}}
\end{equation}
and
\begin{equation}\label{Edef}E = \frac{\nu}{\Omega_\mathrm{o} \ell^2}.\end{equation}
\par
The Taylor-Proudman constraint on rotating flows often holds approximately: overall rotation about the z-axis leads to a tendency toward
z-independence of the flow for large scales or slow
motions. This is useful for nearly steady flow where $(\mathbf{u}\cdot\nabla)\mathbf{u}$ and $\partial\mathbf{u}/\partial t$ in Eq.~\ref{rotNS} are small with respect to
$2\hat{z}\times\mathbf{u}$.      Due to the relatively large size of this apparatus and practical limitations, all of our flows are turbulent.  Thus the Taylor-Proudman theorem is unlikely to hold, though it is expected
that significant anisotropy will remain.  When accelerations are significant, retaining time dependence
but neglecting the nonlinear term  in Eq.~\ref{rotNS} yields dynamics where flow disturbances
can propagate via linear Coriolis-restored
inertial waves~\cite{Greenspan:1968}.  In containers, wave modes arise.  Modes for the interior of a sphere are treated by Greenspan~\cite{Greenspan:1968}.   Zhang \emph{et al.}~\cite{Zhang:2001} provide complete explicit analytical solutions for the full sphere. The modes in a spherical annulus are not known analytically but have been investigated numerically~\cite{Tilgner:1999,Rieutord:2001,Rieutord:1997}.  Inertial waves and related Rossby waves play
a role in the inverse energy cascade to large scales observed in rotating turbulence~\cite{Smith:1999}.  Experiments have shown that inertial modes are important in turbulent spherical Couette flow~\cite{Kelley:2007,Kelley:2010,Schaeffer:2005,Schmitt:2008}, and in more general turbulent flows like rotating grid turbulence~\cite{Bewley:2007}.  We expect that inertial modes have
substantial influence on the character of the flow in rapidly rotating bounded systems for motions with frequencies less than twice the rotation rate.  Above that frequency, no inertial waves
or modes exist.  The question of completeness
of inertial modes is still an open one~\cite{Liao:2009}, so it may not be possible to express arbitrary
motions with frequencies below $\omega = 2\thinspace\Omega_\mathrm{o}$ in terms of inertial mode Fourier components.  However,the nonlinear
interactions of a sea of modes with
$\omega\leq2\thinspace\Omega_\mathrm{o}$ may have a strong signature on a wide range of spatial and temporal scales.
\par
Turbulent flow in a rapidly rotating spherical
annulus with a radius ratio $\eta = r_\mathrm{i}/r_\mathrm{o} =
0.35$ has potential geophysical relevance due to geometrical similarity to
Earth's liquid outer core.  Differential rotation imposed by the boundaries
is at first glance considerably different from the convection, precession, and tidal forcing that may
drive flows in planetary systems.  However, convection in rotating systems tends to set up differential rotation~\cite{Zhang:1992,Zhang:1992a,Aurnou:2001,Aubert:2001,Read:2007,Plaut:2008}, as does the nonlinear interaction of inertial mode shear layers in spherical shells
 driven by precession or tidal deformation~\cite{Tilgner:2007,Morize:2010}.

\par
The study of spherical Couette
flow with rapid outer sphere rotation has been fairly limited.  Experiments and theoretical studies have focused more on the case with the outer
 sphere stationary~\cite{Wulf:1999,Hollerbach:2006,Sisan:2004,Sisan:2004a,Munson:1975,Egbers:1995,Marcus:1987,Marcus:1987a,Beliaev:1978}.  In experiments, this is possibly due to of the difficulty of conducting measurements in the
rotating frame.   Furthermore, there is a complicated dependence of the observed laminar flow on Reynolds number and gap width even when the outer sphere is fixed~\cite{Wulf:1999,Hollerbach:2006}. Nevertheless, some work in
spherical Couette with overall rotation has been carried out.  Hollerbach, Egbers, Futterer, and More~\cite{Hollerbach:2004}
teamed experimental observations with numerical simulations to study Stewartson layer instabilities in
the case of small to modest differential rotation.  They found good agreement between simulation and observed visual patterns in experiments for parameters for $E > 8\times10^{-4}$ and
$0 < Ro < 0.6$.  Experiments by Gertsenshtein, Zhilenko, and Krivonosova~\cite{Gertsenshtein:2001} investigated the transition to turbulence at
some values of $Ro$, especially when $Ro<0$.
Schaeffer and Cardin~\cite{Schaeffer:2005a} studied the
onset of Stewartson layer instabilities at $E$ similar to our experiment at
very low $Ro$ where the first instability happens.
\par
The strongly turbulent behavior of spherical Couette flow has been studied mostly in hydromagnetic apparatus with liquid sodium as a working fluid.  Sisan~\emph{et al.}~\cite{Sisan:2004a} studied instabilities of a turbulent flow of sodium that arise with a sufficiently strong axial magnetic field, with some hydrodynamic measurements to characterize the initial unmagnetized flow.  Kelley \emph{et al.}~\cite{Kelley:2007,Kelley:2010} inferred the flow in a 60~cm rapidly rotating sodium apparatus with a magnetic field sensor array and a dynamically passive applied field.  These experiments demonstrated over-reflectional excitation of inertial modes at comparable Ekman number to ours and $-2<Ro<0$.   Rapidly rotating spherical Couette flow of sodium strongly magnetized by a dipole permanent magnet inner sphere has been studied in the DTS experiment~\cite{Nataf:2006,Schmitt:2008,Nataf:2008} in Grenoble.   Measurements of velocity, magnetic field, and electric potential measurements exposed a number of interesting hydromagnetic states in the turbulent regime~\cite{Nataf:2006,Schmitt:2008,Nataf:2008}.
\par
At present, numerical work on spherical Couette flow has not investigated the portion of the $Ro$, $E$ parameter plane where we find turbulent bi-stability.
However, numerical and theoretical work has demonstrated interesting phenomena with outer sphere rotation.   Stewartson~\cite{Stewartson:1966}, studied
 the case at infinitesimal Rossby number, deriving the form of the free cylindrical shear layer tangent to the inner sphere equator that still bears his name.  A systematic three-dimensional numerical study of instabilities of the Stewartson was undertaken by
Hollerbach~\cite{Hollerbach:2003}.
Schaeffer and Cardin~\cite{Schaeffer:2005,Schaeffer:2005a} used depth-averaged equations of motion coupled to
realistic Ekman pumping to conduct quasi-geostrophic simulations at low Ekman number for a split-sphere geometry similar to spherical Couette flow. Schaeffer and Cardin~\cite{Schaeffer:2005a} showed good agreement between experiment and simulation, predicting Stewartson layer
instabilities correctly for very low Rossby number.  Quasi-geostrophic simulations~\cite{Schaeffer:2005}  found Rossby wave turbulence at higher $Ro$.
\par
Our experiments can not generally achieve $Ro<0.05$ at experimentally accessible Ekman number because of limitations
on motor minimum speeds.  The lowest possible Rossby number increases if we rotate the system more slowly to raise $E$.   It is impractical to reduce the Rossby number or increase the Ekman number enough to match the parameters of known numerical simulations.
\par
Guervilly and Cardin~\cite{Guervilly:2010} performed simulations of spherical Couette flow with outer rotation in an investigation of magnetic dynamo action.  They performed fully three dimensional numerical simulations for Ekman number higher than $E=10^{-4}$ while achieving
Rossby number matching some presented here.   The definition of the Rossby number used by Guervilly and Cardin~\cite{Guervilly:2010} is different, $Ro_\mathrm{GC} = \eta\Delta\Omega/\Omega_\mathrm{o} = \eta Ro$.
They report an $m=2$ Rossby type wave for $Ro\sim2.9$ ($Ro_\mathrm{GC}=1$), above our first bistable range.  The existence and azimuthal wavenumber of this wave is partially consistent with our observations at that Rossby number.
However, we observe strong turbulence, additional waves, and the bi-stable behavior that is the focus of this paper.  These have not been previously reported in spherical Couette flow.
\par
Turbulent flow transitions and multiple stability in very high Reynolds number flows have been reported in a
number of systems.
Several examples exist in geophysics and astrophysics.
The dynamo generated magnetic fields of the Earth and Sun reverse polarity.  Ocean currents,
namely the Kuroshio current in the North Pacific near Japan and the Gulf Stream, both exhibit bi-stability in
 meander patterns.~\cite{Schmeits:2001}  Polar vortices in Earth's stratosphere are bounded by resilient transport
 barriers much of the time, mixing to higher latitudes only intermittently.~\cite{Bowman:1996,Rypina:2007,Haynes:2005}  This process
 is important to ozone depletion in polar regions and may have some similarities to the dynamics we observe.
\par
A number of laboratory flows are known to exhibit multi-stability and hysteresis. The mean circulation in
turbulent thermal convection cells has been observed to switch
direction abruptly.~\cite{Sreenivasan:2002}  Hysteresis in the
large scale flow states has been seen in surface waves excited by
turbulent swirling flows in a Taylor Couette geometry with a free
surface.~\cite{Mujica:2006} Von  K\'{a}rm\'{a}n flow in a cylinder between
two independently rotating impellers has  exhibited
multi-stability and hysteresis of the mean flow despite extremely high fluctuation
levels.~\cite{Ravelet:2004,Torre:2007,Cortet:2010} Magnetohydrodynamic experiments in the von  K\'{a}rm\'{a}n geometry have succeeded in
producing dynamos that show reversals of the generated magnetic
field.~\cite{Monchaux:2009,Berhanu:2007}
The L-H transition in turbulent tokamak plasma confinement devices involves the formation of a wave-driven zonal flow transport barrier that greatly
enhances confinement of the plasma.~\cite{Connor:2000,Guzdar:2001}  However, this barrier eventually breaks down in a burst that can damage the confinement device.~\cite{Kleva:2007}
Understanding this particular form of turbulent multiple
stability and its control is important issue in sustained confinement of fusion plasmas.
\par
Spherical Couette flow is a dynamically rich system in both laminar and turbulent regimes.  Our apparatus,
initially designed, constructed, and eventually destined for
magnetohydrodynamic experiments in molten sodium metal,
presents a unique opportunity to measure the properties of
hydrodynamic turbulence in this geometry, including transition phenomena between multiple turbulent states.  We operate in a novel
region of parameter space, simultaneously achieving moderately high Rossby number and low Ekman number.  This regime can currently only be directly accessed by experiments and naturally occurring flows, and
has not been the focus of previous purely hydrodynamic studies.  Furthermore, we are able to make quantitative measurements in the rotating frame, something that can be quite difficult in smaller apparatus.
\section{\label{expt}Apparatus}
\par
\begin{figure}[ht]
 \includegraphics[width=7cm]{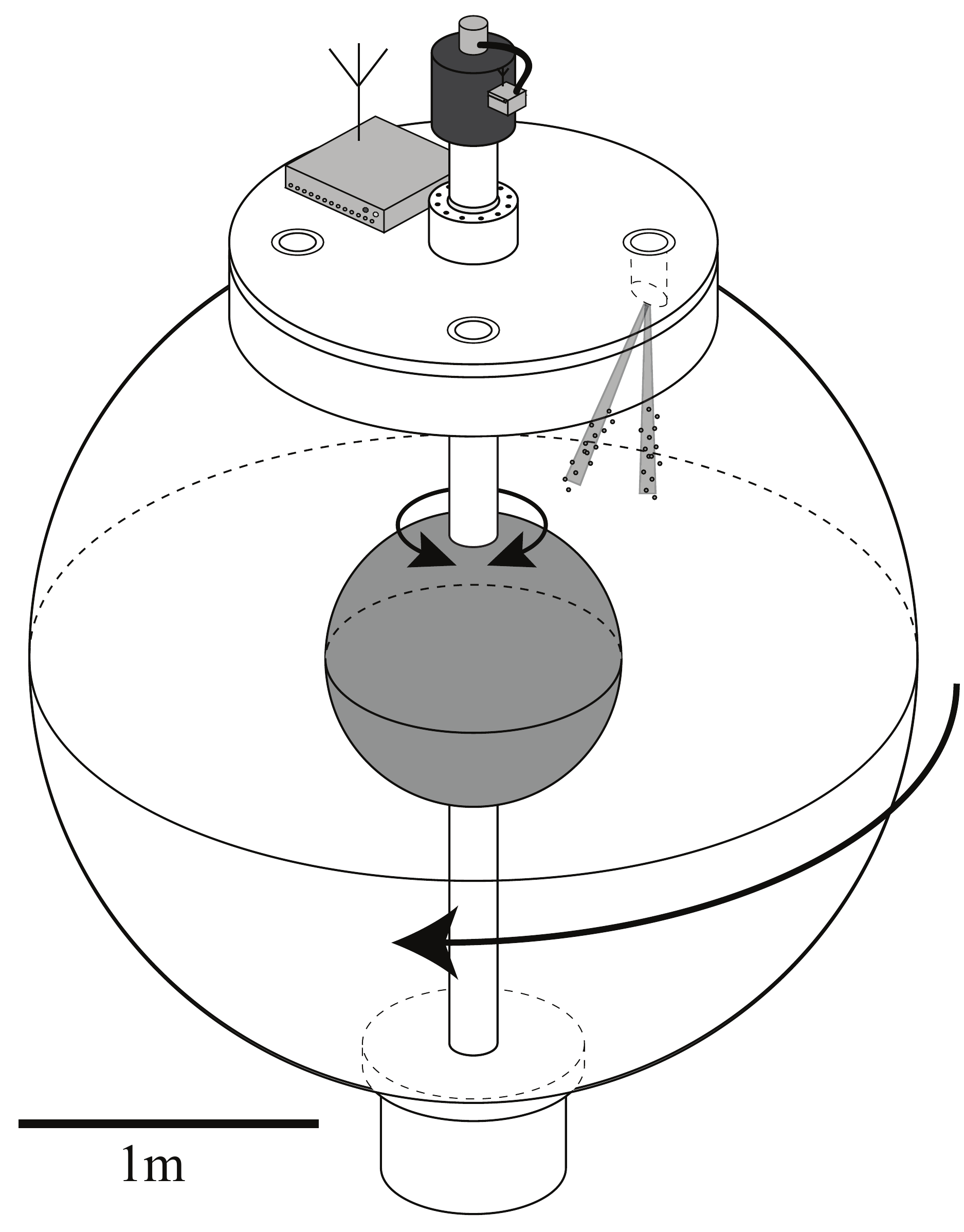} 
 \caption{\label{experiment}{A schematic of the apparatus showing the inner and outer sphere and locations of
 measurement ports in the vessel top lid at 60~cm cylindrical radius $(1.18~r_\mathrm{i})$.  Data from sensors in the ports are acquired by
 instrumentation, including an acquisition computer, bolted to the rotating lid and wirelessly transferred to the lab frame.  Also shown
 is the wireless torque sensor on the inner shaft.}}
 \end{figure}

The three meter apparatus allows independent rotation
of the inner and outer spheres.
Instrumentation in the rotating frame allows measurements of velocity, wall shear and
pressure, as well as the torques required to maintain the
boundary speeds.  Fig.~\ref{experiment} is a schematic sketch of the apparatus.
  The stainless steel outer vessel has an inner diameter of
$2.92\pm0.005$~m and is 2.54~cm thick. It is
mounted on a pair of spherical roller bearings held by a frame.    The vessel top lid is installed in a 1.5~m diameter cylindrical flanged opening,
and the inside lid surface is curved to complete the outer
spherical boundary.  The lid has four 13~cm diameter instrumentation ports
centered at 60~cm cylindrical radius.  Due to design constraints aimed at safe operation
with liquid sodium metal as the working fluid, these four ports are the only penetrations
through the outer boundary, and so are the only location from which we may make direct flow measurements.
 Port inserts hold
measurement probes nearly flush with the inner surface of the
outer sphere.
\par
 The inner sphere has a diameter of $1.02\pm0.005$~m  and is
supported on a 16.8~cm diameter shaft held coaxial with the outer
shell by bearings at the bottom of the outer sphere and in the top
lid.  The inner sphere is driven from a 250~kW electrical motor through a calibrated
Futek model TFF600 torsional load cell.  The measured torque includes the torque from a pair of lip seals
which add some confounding error.  The outer sphere is driven by a 250~kW induction motor
mounted to the support frame. A timing belt reduction drive with a 25:3 ratio couples the outer sphere motor to a
toothed pulley on the lid rim.
\par
 Motor speeds are controlled to better than 0.2\% by variable frequency drives, and optical encoders monitor the inner
and outer sphere speeds.  The drives estimate the motor torque from electrical current
measurements and the torque estimate supplied by the inner motor drive agrees well with the
calibrated strain-gauge torque sensor at motor speeds above about 2~Hz.  This supports the use of the outer motor drive's reported torque as a reasonable estimate
 of the total torque required to drive the outer sphere, provided that the outer sphere angular speed is above about 0.24~Hz, as it
is for the data presented here.  As the outer sphere speed is lowered below 0.24~Hz, the outer drive's reported torque becomes increasingly dominated by motor magnetizing current.  The torque exerted by aerodynamic and bearing drag on the
outer sphere is typically larger than the working fluid's contribution.  However, this drag is repeatable and
can be subtracted off by measuring the torque demanded with no differential rotation.

\par A rotating computer acquires data from sensors in the instrumentation ports at a sampling rate of 512~Hz, recording data on
a lab frame computer using a wireless ethernet connection.  Sensors include a
Dantec model 55R46 flush mount shear stress sensor driven by a TSI
model 1750 constant temperature anemometer and three Kistler model
211B5 pressure transducers. The three pressure transducers are installed in three ports $90^\circ$ apart on the 60~cm radius
port circle. A
thermocouple is used to monitor fluid temperature.
\par A Met-Flow UVP-DUO pulsed Doppler ultrasound velocimeter is
mounted in the rotating frame and paired with Signal Processing
transducers, also communicating with the lab frame using wireless ethernet.
Some velocity data in this paper was
acquired with a Met-Flow UVP-X1-PS ultrasound unit in the lab frame using a
resonant transformer arrangement to couple the signal into the rotating frame.
To scatter ultrasound, the flow is seeded with
$150~\mathrm{\mu m}-250~\mathrm{\mu m}$ polystyrene particles with
nominal density of $1.05~\mathrm{g/cm^3}$.   Limitations
on the product of maximum measurable velocity and measurement
depth constrain the velocity measurements in this paper to be very
close to the wall, typically no more than 10~cm from the transducer face, which is
either flush mounted with the wall or intruding no more than 10~cm into the flow.  Intrusive
transducers do not seem to measurably change the torque dynamics.  When possible, we measure upstream to avoid
directly measuring the turbulent wake shed from an intrusive mounting scheme.

 \section{\label{Gsec}Torque Measurements}
 \par
The torques required to maintain constant speeds of the boundaries of in a rotating flow provide a global picture of angular momentum transport
 and power dissipation.  Experimental torque measurements on wall-driven flows have been largely
 confined to the flow between concentric cylinders\cite{Lathrop:1992,Lewis:1999,Wendt:1933,Paoletti:2011,Gils:2011,Ravelet:2010}.  The torque required to drive the flow between two spheres has
received less attention, with previous results in the hydrodynamic turbulent regime seemingly limited to outer-sphere stationary measurements of liquid sodium~\cite{Sisan:2004}.
\begin{figure}[ht]
 \includegraphics[width=8.6cm]{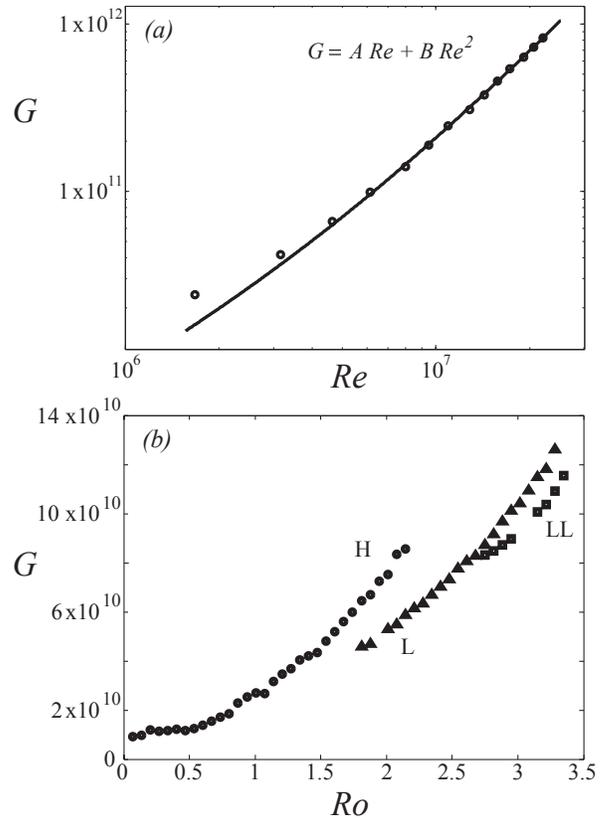}
 \caption{\label{GRe}{$(a)$ The dimensionless torque $G$ vs. the Reynolds number $Re$ for the case
 $\Omega_\mathrm{o} = 0$, stationary outer sphere, with a fit to
 $G = A Re + B Re^2$ with $A = 4.49\times10^{4}$ and $B = 0.05$.   $(b)$ Mean value of the dimensionless torque $G$ vs. $Ro$ at $E = 2.1\times10^{-7}$.  $Ro$ is varied by increasing the inner sphere speed
 in steps of $Ro = 0.067$, waiting 450 rotations per step.  Three curves are differentiated by circles, triangles, and
 squares.  In the ranges of $Ro$ where the symbols overlap, flow exhibits bistable behavior.  H and L denote the torque curves of the ``high torque" and ``low torque" states.
 There is a second bistable regime starting around $Ro = 2.75$, with LL labeling the lower torque state.   }}
 \end{figure}
Following Lathrop \emph{et al.}~\cite{Lathrop:1992b}, we define the dimensionless
torque $G$ on the inner sphere,

\begin{equation}\label{Gdef}
G = \frac{T}{\rho\nu^2r_\mathrm{i}},
\end{equation}
 where $\rho$ is the fluid density, $\nu$ is the kinematic viscosity, $r_\mathrm{i}$ is the inner sphere
radius and $T$ is the dimensional torque.  The torque $G$ as a function of Reynolds number $Re$, here defined as \begin{equation}\label{Redef}Re =
\frac{(\Omega_\mathrm{i}-\Omega_\mathrm{o}) \ell^2}{\nu} ,\end{equation} is bounded above
by $G\propto Re^2$ from dimensional arguments. We fit the measured torque with the
outer sphere stationary to
 $G \propto Re^2$ with a correction for linearly increasing
seal drag.   Data and fit with the outer sphere stationary are shown
shown in Fig.~\ref{GRe}$(a)$ for the range of $Re$ accessible
in this experiment. At the lowest rotation rates, the measurement is significantly
confounded by the bearing and seal torque.

Figure~\ref{GRe}$(a)$ shows that $G$ monotonically
increases as $Re$ is increased with the outer sphere stationary. This is not necessarily the case
when we impart global rotation, provided that we do not hold $Ro$ constant.  When we hold the outer sphere
speed constant (constant Ekman number) and super-rotate the inner sphere ($Ro>0$) as in Fig.~\ref{GRe}$(b)$, we find that the mean torque on the inner sphere is non-monotonic as the dimensionless differential speed, $Ro$, and the Reynolds number, $Re=Ro/E$ are concurrently increased.   As $Ro$ is increased, the flow undergoes several
transitions to significantly different turbulent flow states.  In ranges of $Ro$ in Fig.~\ref{GRe}$(b)$
where the torque data are presented in overlapping branches with different symbols,
the flow exhibits bistable behavior with intermittent transitions between adjacent flow states.   In these ranges of $Ro$, we plot the mean torque of each state, conditioned on state.

   \begin{figure}[ht]
 \includegraphics[width=8.6cm]{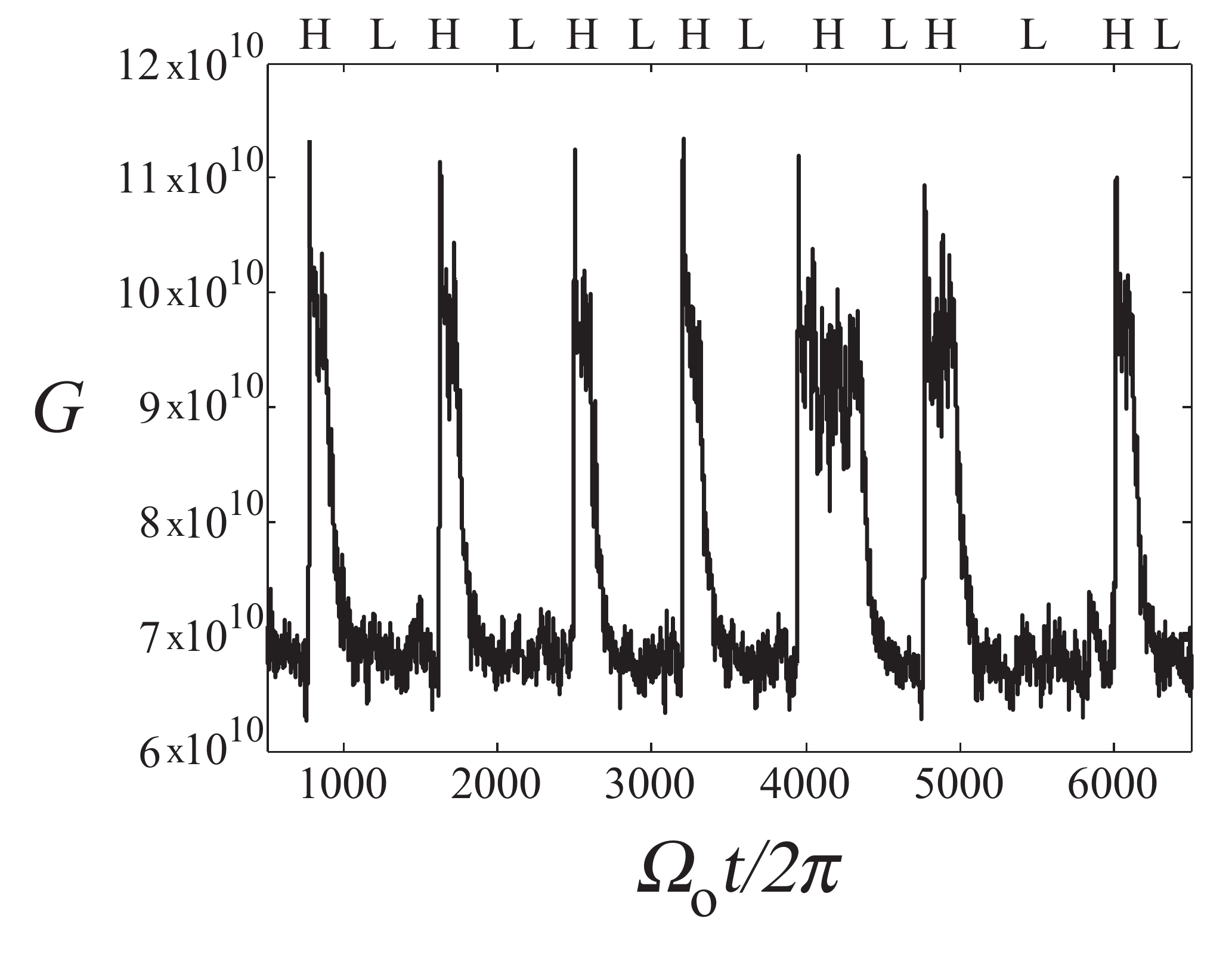}
 \caption{\label{Gtime}{Time series of $G$ at fixed $Ro = 2.13$ and $E=2.1\times10^{-7}$, with time made dimensionless
  by the outer sphere rotation period.
  The raw torque signal has been numerically low pass filtered ($f_c
  = 0.05$~Hz, 15 rotations of the outer sphere.).}}
 \end{figure}

  The state
switching means that the transitions are not hysteretic, and long time mean torque through the switching range decreases with increasing $Ro$.  We choose to plot the conditioned torque branches here
instead of the long-time mean to emphasize our observation that these states seem the same as those present alone at higher and lower $Ro$.  We will focus on the first bistable
regime in this paper, between $Ro=1.8$ and $Ro=2.75$,  and not the second bistable regime with lower state ''LL" that begins near $Ro = 2.75$.  This second bistable regime has several
similarities to the first, but we will not discuss it in detail in this paper.
\par
A representative time series of the torque in the first bistable regime is shown in Fig.~\ref{Gtime}, with $E=2.1\times10^{-7}$ and $Ro = 2.13$.
 The onset of the high torque state is abrupt, taking on the
 order of 10 rotations of the outer sphere.  The torque
 overshoots the high state mean value at high torque
 onset by 10-15\%.

 The
 end of the high torque state exhibits a slow decay of the torque to the
 low torque value, approximately exponential with a time constant of 40 rotations of the outer sphere.
 In addition to the full transitions between the two
 torque levels, there are ``excursions" where the torque decays toward the low mean value or rises toward the high mean value without
 fully reaching the other state.
 \par
\begin{figure}[ht]
 \includegraphics[width=8.6cm]{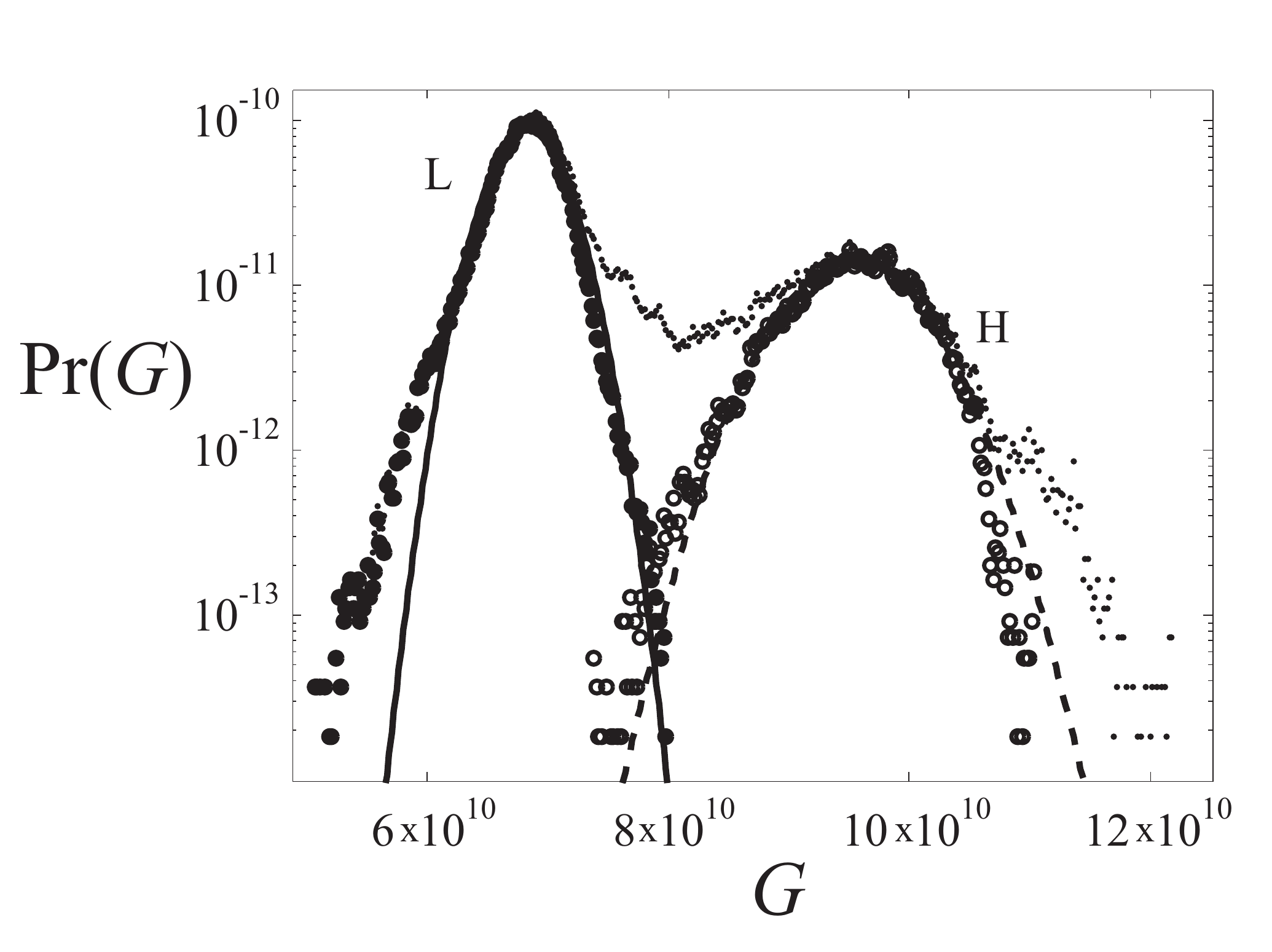}
 \caption{\label{GPDF}{Probability density of the dimensionless torque at $Ro=2.13, E = 2.1\times10^{-7}$.
 The full, unconditioned distribution is denoted by small points.  Solid circles denote conditioning on low torque state, and open circles on the high, with
 Gaussian solid and dashed curves for the low and the high respectively.  The mean and standard deviation of the low torque state data are $\langle
 G\rangle = 6.82\times10^{10}$ and $\sigma_{G} = 2.74\times10^9$.
 In the high torque state they are $\langle
 G\rangle = 9.53\times10^{10}$ and $\sigma_{G} = 5.02\times10^9$.  The data were low pass filtered at $f_c = 0.5$~Hz to remove high frequency noise caused by mechanical vibration.    }}
 \end{figure}

 The qualitative bi-stability in Fig.~\ref{Gtime} can be expressed
 quantitatively in the bimodal probability distribution of the torque shown in Fig.~\ref{GPDF}.  The division of data
 into high torque and low torque states was done manually so as to exclude the transition regions.  The resulting individual distributions of $G$ for the high and low states based on this conditioning are shown in Fig.~\ref{GPDF}, as well as the full distribution.  There is a small region of overlap in the H and L state individual distributions due to the difference drawn between ``transitions" and ``excursions."  The same manual division in states is used throughout the paper to condition other data on state.  In Fig.~\ref{GPDF}, the high state mean torque is 1.4 times that in the low state.  The torque fluctuations in the high
 torque state are considerably higher than in the low; the standard
 deviation of the high state torque is 1.8 times that in the low
 state.  The low frequency fluid fluctuations responsible for this are also observed in the velocity and wall shear.

 In both Fig.~\ref{Gtime} and Fig.~\ref{GPDF} the torque data has
 been numerically low pass filtered.  In Fig.~\ref{Gtime} the cutoff frequency is 0.05~Hz.  In Fig.~\ref{GPDF}, the
 filter cutoff frequency was chosen to be 0.5~Hz, where the fluid torque power spectrum appears to cross the mechanical vibration noise
 floor.  In this way we retain the fastest measurable hydrodynamically relevant fluctuations.
 \par
\begin{figure}
 \includegraphics[width = 8.6cm]{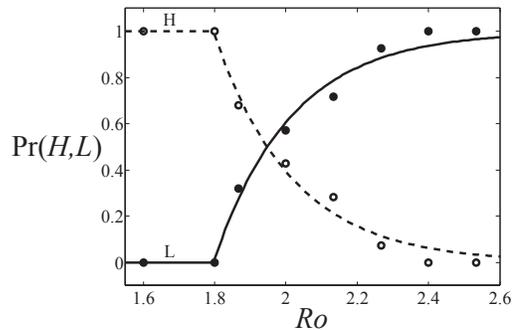}
 \caption{\label{stateprob}{Probability that the flow was in the
  high torque (open circles) or low torque (closed circles) state as a
 function of $Ro$ over the first bistable range with fixed $E =
 2.1\times10^{-7}$.  The dashed and solid lines are fits to the exponential form
 of Eq.~(\ref{fitform}).}}
\end{figure}

 \begin{table}[hb]
 \caption{\label{rotinterval} Statistics of the interval between high torque onsets.  $\Delta t^{\prime}_\mathrm{H}$ is the time
 interval between two subsequent high torque onsets made dimensionless by $\Omega_\mathrm{o}/2\pi$, so the time interval is measured
 in outer sphere rotations.  $Ro =2.13$, $E = 2.1\times10^{-7}$. }
 \begin{ruledtabular}
 \begin{tabular}{c c c c}
 $\langle \Delta t^{\prime}_\mathrm{H}\rangle$ & $\sigma_{\Delta t^{\prime}_\mathrm{H}}$ & Max($\Delta t^{\prime}_\mathrm{H}$) & Min($\Delta
 t^{\prime}_\mathrm{H}$)\\[3pt]
 \hline
 717 & 313 & 1917 & 390\\
 \end{tabular}
 \end{ruledtabular}
 \end{table}
 The interval between transitions is somewhat irregular.  Over 45 transitions at the parameters in Fig.~\ref{Gtime}, we
  observe the statistics shown in Table~\ref{rotinterval}.
 We also observe that the probability that the system is in one state or the other depends on $Ro$.
 Above a threshold value of $Ro$, we begin to observe state transitions to the low state, and the high torque state becomes less likely as $Ro$ increases.   Fig.~\ref{stateprob}
 shows the probability that the system is in the high or low
 torque state for 4000 rotations of the outer sphere across the first bistable range of $Ro$.
 At values of $Ro$ where transitions were not observed for more than
 4000 rotations of the outer sphere, a probability of one or zero was assigned.   The
 probability that the system was in the low state was fit to
 \begin{equation}\label{fitform}
 \mathrm{Pr(}L\mathrm{)} = \left\{
 \begin{array}{lr}
    0 & : Ro< Ro_\mathrm{c}\\
    1-exp(-\gamma(\frac{Ro-Ro_\mathrm{c}}{Ro_\mathrm{c}}))& :
    Ro > Ro_\mathrm{c}
    \end{array}
    \right.
    \end{equation}
 with $\gamma = 8.25$ and $Ro_\mathrm{c} = 1.80$.  The
high torque
 probability
 is given by Pr($H$) = 1-Pr($L$).  The physical implication inherent in the exponential form is that there is
 no upper threshold where the high torque state becomes impossible.  Instead, it only becomes less
 likely as $Ro$ is increased.  However, the lower threshold for state transitions is well defined at $Ro_\mathrm{c}$.

 \section{\label{L}Fluid Angular Momentum}
 The torque measurements presented so far only considered  the torque on the inner
 sphere.  To see the complete picture of the angular momentum transport
 in the system, we examine the torque on both
 boundaries. The torque on the outer sphere is reported by the motor drive, though with less precision than the inner torque sensor measurement.
 \par
\begin{figure}[ht]
 \includegraphics[width=8.6cm]{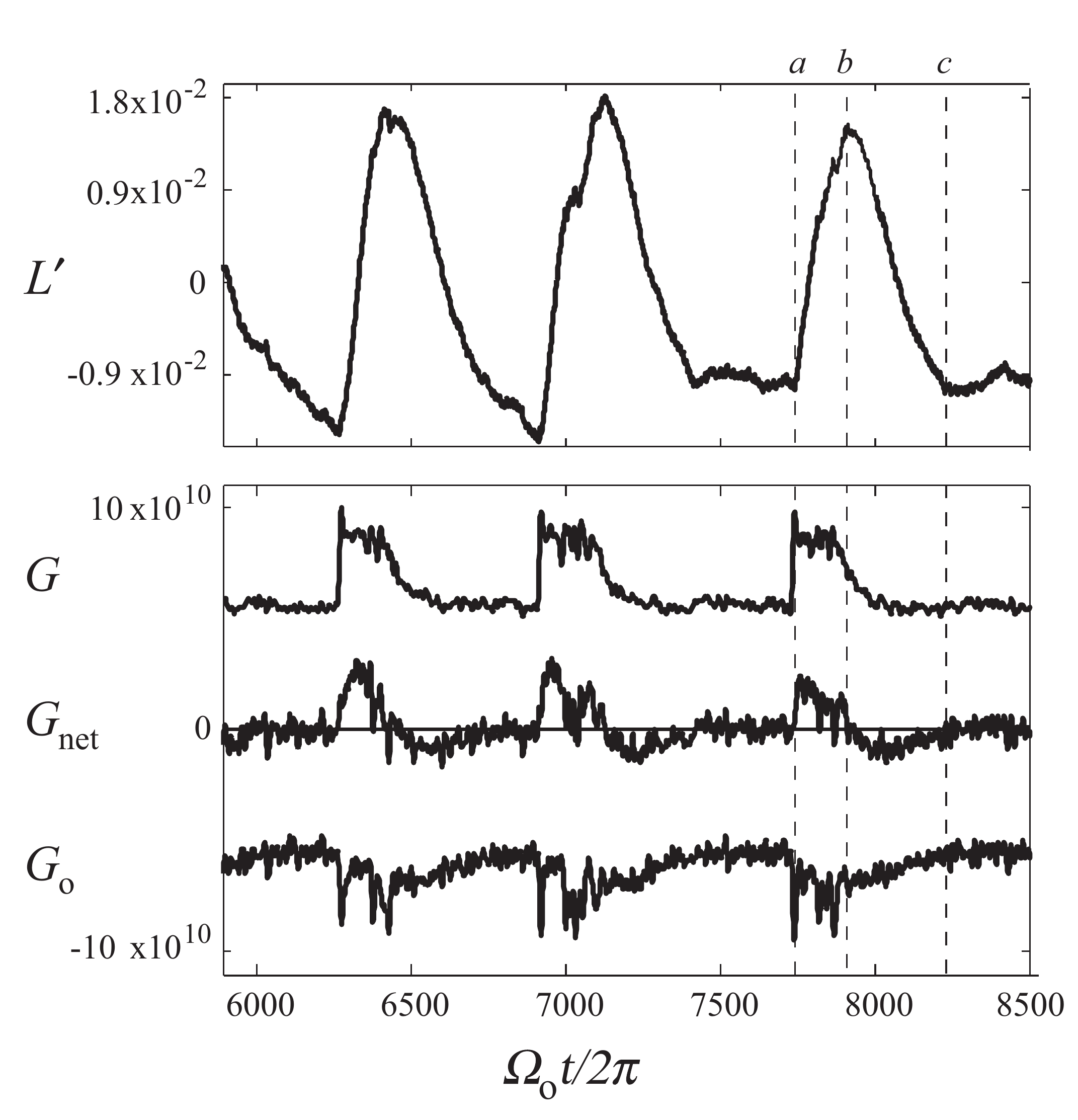}
 \caption{\label{GnL}{The angular momentum, inner torque, outer torque, and net torque.  $Ro = 2.13$, $E=2.1\times10^{-7}$.  The upper plot shows the
 dimensionless angular momentum $L^{\prime}$ as defined in Eq.~(\ref{ndL}).  The lower plot shows the inner torque $G$, the outer torque
 $G_\mathrm{o}$, and their sum $G_\mathrm{net}$.  The bearing and aerodynamic drag on the
 outer sphere have been subtracted off, and the torques have been
 low pass filtered as in Fig.~\ref{Gtime}.
 }}
 \end{figure}

 The fluid filling the gap cannot undergo angular acceleration or deceleration for arbitrarily long times with the boundaries rotating at constant speeds.  Therefore,
 the time averaged net torque on the fluid must be zero, and the boundary torques must be equal and opposite, provided averaging is done over sufficiently long times.
 In the bistable regime, however, the net torque, shown in Fig.~\ref{GnL}, reveals
 long periods of angular acceleration and deceleration interspersed with plateaus where the angular momentum of
 the fluid remains nearly constant.  The torques on the inner and outer boundaries only balance on averaging over many state transitions.  We define the net torque on the fluid as
 \begin{equation}\label{netdef}
 T_\mathrm{net} = T+T_\mathrm{o},
 \end{equation}
 where $T$ and $T_\mathrm{o}$ are the torques about the axis that the fluid exerts on the inner and outer spheres respectively. $T_\mathrm{o}$ is always negative when $Ro>0$, as the fluid would tend to speed up the outer sphere rather than resist its motion.
 The total torque on the outer sphere as reported by the outer sphere motor drive, $T_\mathrm{o, meas}$, is largely due to drag from the outer sphere bearings and air around the outer sphere.  However, it is reproducible, and we define the torque that the interior fluid exerts on the outer sphere $T_\mathrm{o}$ in Eq.~\ref{netdef}, as
 \begin{equation}\label{todef}
 T_\mathrm{o} = T_\mathrm{o, meas}-T_\mathrm{drag}.
 \end{equation}
  The drag, $T_\mathrm{drag}$ is the outer drive torque required to keep the experiment in solid body rotation (with inner locked to outer) at the same outer sphere angular speed.  In the $Ro>0$ flow states, some of the total torque required to maintain the speed of the outer shell against bearing and aerodynamic drag is supplied by the inner sphere via the fluid, and this is easily detected.

\par The net torque is shown in Fig.~\ref{GnL} along with the separate inner and outer
torques with the steady outer sphere bearing and aerodynamic drag subtracted.  The dimensionless torques are defined as in Eq.~\ref{Gdef}, for example
$G_\mathrm{net} = T_\mathrm{net}/\rho\nu^2 r_\mathrm{i}$.
Up to a constant of integration, we can calculate the fluid angular momentum $L(t)$ about the rotation axis
from the net torque,
\begin{equation} \label{Ldef} L(t) = \displaystyle\int^t_0T_\mathrm{net}(t^{\prime})dt^{\prime}. \end{equation}
Due to measurement limitations, we cannot integrate the torques
from motor startup to fix the constant of integration.
 Instead, we set the initial value of $L$ to zero at an arbitrary time and make the resulting quantity
 dimensionless by dividing it by the angular momentum the fluid would have if it were in solid body rotation with the
 outer sphere,
 \begin{equation}\label{ndL}L^{\prime} = \frac{L}{I_\mathrm{fluid}\Omega_\mathrm{o}}.\end{equation}
 The moment of inertia of the fluid filling the gap is
\begin{equation}\label{Iflu} I_\mathrm{fluid} = \frac{8\pi\rho}{15}(r_\mathrm{o}^5-r_\mathrm{i}^5),\end{equation} with a value of $(1.14\pm0.02)\times10^4~\mathrm{kg~m^2}$.

At the onset of the high torque state, as shown in Fig.~\ref{GnL}$(a)$, there is some prompt
response of the outer sphere torque, indicating a certain amount of increased
angular momentum transport. However, the increase in the torque on the
outer sphere is insufficient to fully oppose the increased inner sphere torque.
At this point, the net torque becomes steadily positive, and the
fluid accelerates.  As this happens, the fluid torque $G_\mathrm{o}$ on the
outer sphere tends to slowly decrease in magnitude, though with large fluctuations. Eventually the inner sphere torque $G$ starts the slower
transition to the low torque state.  At a point during the high to low transition, as shown in Fig.~\ref{GnL}$(b)$, the net torque becomes
negative and the total angular
momentum starts to decrease. The torque on the outer sphere continues
a slow decay toward a value opposite and equal to that on
the inner sphere, occasionally reaching a net torque fluctuating about zero as in times after Fig.~\ref{GnL}$(c)$.

 \par When both spheres rotate
the angular momentum fluctuations in the range of $Ro$ considered here may indicate a slow ``store and release" process where long lasting imbalances
in the torques on the inner and outer spheres lead to fluid spin up and spin down.
We note that the average magnitude of torque is similar between the outer stationary and the outer
rotating cases for the same Reynolds number in Fig.~\ref{GRe}(a) and Fig.~\ref{GRe}(b).
In both cases the mean torque on the inner sphere is between $G=10^{10}$ and $G=10^{11}$.
Although the dynamics of the angular momentum
transport seem quite different, the magnitude of the transport
has not changed drastically.

 \section{\label{flow}Mean Flow Measurements}
\begin{figure}
 \includegraphics[width=8.6cm]{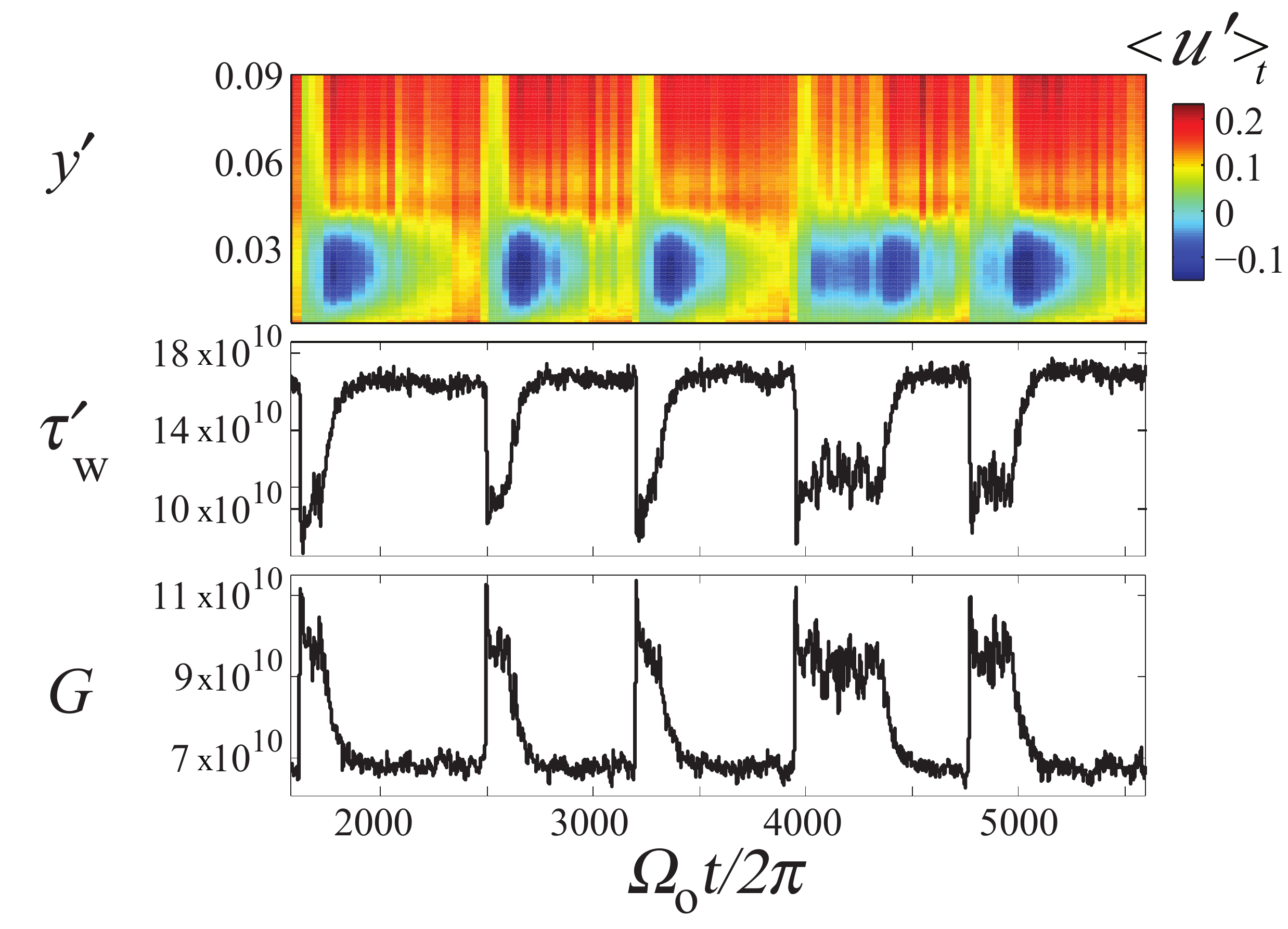}
 \caption{\label{UtauG}{Simultaneous time series of velocity, wall shear stress, and torque.  A space-time diagram of the low pass filtered velocity $u^{\prime}$ (defined in Eq.~\ref{uprimedef}) is shown at the top.
  This measurement is dominated by the azimuthal velocity $u_\mathrm{\phi}$.  The velocity $u^{\prime}$ is made dimensionless by the outer sphere
 tangential velocity, and so can be interpreted as a locally measured Rossby number.
  The  dimensionless wall shear stress
 $\tau^\prime_\mathrm{w} = 4\pi r_\mathrm{o}^3 \tau_\mathrm{w} /(\rho\nu^2r_\mathrm{i})$ is shown in the middle, and the dimensionless torque $G$ is shown at the bottom.  The wall shear stress and torque have been low pass filtered with $f_c = 0.05Hz$ as before, which is comparable to the time averaging of the velocimetry.}}
 \end{figure}

In the bistable regime, all measured flow
quantities undergo transition along with
the inner torque.  Fig.~\ref{UtauG} shows that the time averaged dimensionless wall shear $\tau_\mathrm{w}^\prime$ and measured
dimensionless azimuthal velocity $u^\prime$ decrease sharply the onset
of the high torque state.
\par In Fig.~\ref{UtauG}, the velocity was measured in a shallow range near the surface of the outer sphere at $23.5^\circ$ colatitude (60~cm cylindrical radius).  This location is about 9~cm outside a vertical cylinder tangent to the inner sphere equator (called simply the ``tangent cylinder" from here on).  In  Fig.~\ref{UtauG}, the transducer beam was in a plane normal to the cylindrical radius at the port and was inclined $23.5^\circ$ from pointing straight down ($23.5^\circ$ inclined from negative $\hat{z}$).  Since the transducer responds only to the velocity component along the beam axis, this orientation made it sensitive to the cylindrical radial, azimuthal,
and vertical velocity components, $u_\mathrm{s}$,
$u_{\mathrm{\phi}}$, and $u_\mathrm{z}$.
\par
However, by using a remotely controlled rotatable mount, we determined that the time-averaged measured velocity is dominated by the azimuthal component, $u_\mathrm{\phi}$.  When the transducer is rotated $90^\circ$ from the usual orientation, making it least sensitive to $u_\mathrm{\phi}$, the mean velocity was about 10\% of that
seen when it maximally responds to the azimuthal flow.  This indicates a meridional circulation of some importance.  But it also means that the meridional circulation provides a small contribution to the mean measured velocity when the transducer is oriented to be most sensitive to azimuthal flow. This is the case in Fig.~\ref{UtauG}, and so we treat the measured mean velocity as entirely azimuthal and correct for the vertical inclination, so that:
\begin{equation}\label{uphi}\langle u_{\mathrm{\phi}}\rangle_t = \frac{\langle u_\mathrm{meas}\rangle_t}{\sin{23.5^{\circ}}}.
\end{equation}
We then divide by the outer sphere equatorial tangential velocity to make $\langle u_{\mathrm{\phi}}\rangle_t$ dimensionless:
\begin{equation}\label{uprimedef}\langle u^{\prime}\rangle_t = \frac{\langle u_\mathrm{\phi}\rangle_t}{\Omega_\mathrm{o} r_\mathrm{o}}.
\end{equation}  This non-dimensionalization means that $u^\prime$ can be interpreted as a local Rossby number.
\par  The wall shear stress sensor was calibrated against the
measured torque $T$ on the inner sphere with the outer
sphere stationary.   We assumed that the mean wall shear $\tau_\mathrm{w}$ on the
outer sphere was
\begin{equation}\label{wsmean}\tau_w = \frac{T}{4\pi r_o^3
\cos{23.5^\circ}}. \end{equation} We fit the
bridge voltage $V$ and mean wall shear
calculated from the measured torque by Eq.~(\ref{wsmean}) to
\begin{equation} \label{wsfit}V^2 = A\tau_\mathrm{w}^{2/3} + B\tau_\mathrm{w}^{1/3}
+ C,\end{equation} as was done by Lathrop~\emph{et al.}~\cite{Lathrop:1992} We then used the
calibration coefficients $A$,$B$ and $C$ to calculate the wall shear stress $\tau_\mathrm{w}$ from the measured
bridge voltage for all subsequent data with the outer sphere rotating.

\par
\begin{figure}
 \includegraphics[width=8.6cm]{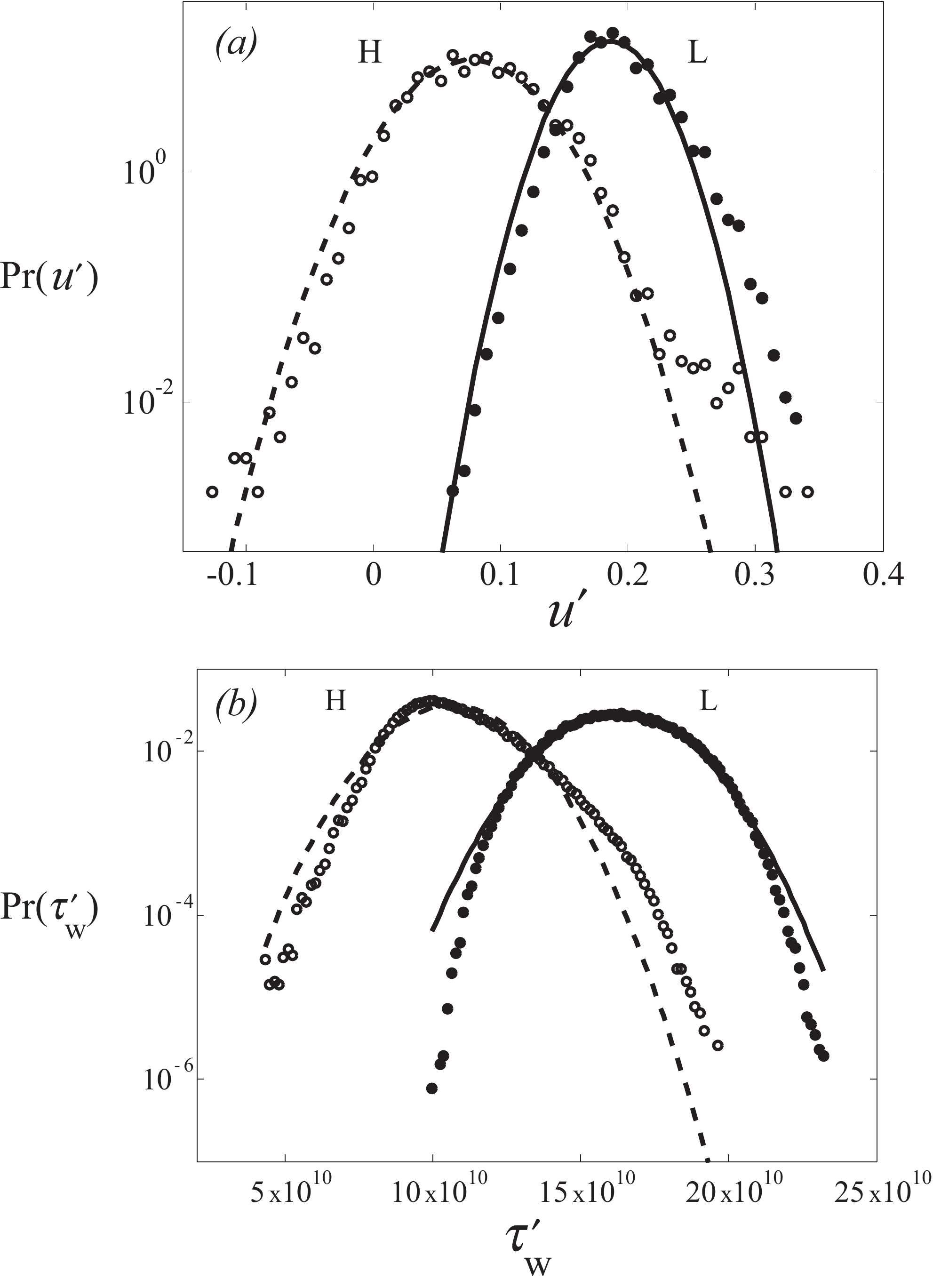}
 \caption{\label{UPDF}{$(a)$~The probability density of the dimensionless velocity conditioned on the torque state, with solid circles denoting
 the low torque state and open circles denoting the high.  Dashed and solid lines are Gaussian.  Standard deviations are $\sigma_{u^{\prime}} = 0.043$ for the high state,  $\sigma_{u^{\prime}} = 0.029$ for the low state.  The
 high state mean is $\langle u^{\prime}\rangle = 0.076$, and the low state mean is $\langle u^{\prime}\rangle = 0.186$.  The means are dominated by azimuthal velocity, though the transducers are equally sensitive to $u_\mathrm{z}$ and $u_\mathrm{\phi}$ in this measurement. $(b)$~The probability density of the dimensionless wall shear conditioned on torque.
 Solid circles again denote the low torque state, and open circles the high, with solid and dashed Gaussian curves. The mean and standard deviation in the low torque state are $\langle
 \tau_\mathrm{w}^{\prime}\rangle = 1.63\times10^{11}$ and $\sigma_{\tau_\mathrm{w}^{\prime}} = 1.81\times10^{10}$.
 In the high torque state they are $\langle
 \tau_\mathrm{w}^{\prime}\rangle = 1.06\times10^{11}$ and $\sigma_{\tau_\mathrm{w}^{\prime}} = 1.71\times10^{10}$.   }}
 \end{figure}

 In the
torque switching regimes, the slow fluctuations in
wall shear and the mean azimuthal velocity shown in Fig.~\ref{UtauG} are both strongly anti-correlated with the inner sphere torque.  The simultaneous wall shear and torque data suggest that the two different states have different latitudinal distribution of the shear stress on the outer sphere.     Fig.~\ref{GnL} shows that the torque the fluid exerts on the
whole outer sphere is indeed somewhat greater in the high torque state.    However, the mean shear stress at
the measurement location in the high torque state is 65\% of that measured in the low torque state (see also Fig.~\ref{UPDF}$(b)$), implying a shear stress concentration at high latitudes in the low torque state relative to the distribution of shear stress in the high torque state.
\par
These shear and velocity measurements suggest a fast central zonal flow in the low torque state that ceases suddenly at the high torque onset.
When the fluid around, above, and below the inner sphere
is circulating faster in the low torque state, there is less drag on the inner sphere.  The unusual
aspect of this is that it takes less torque and less power input to maintain
the faster circulation.  This could indicate a transport barrier to
energy and angular momentum in the low torque state.  This change in transport is also important in interpreting the
observation that the total angular momentum is often decreasing in the low torque state (See Fig.~\ref{GnL}, line $(b)$), while the measured circulation is increasing or steady. This is only possible
with a change in shape of the angular momentum profile as a function of cylindrical radius.  An angular momentum transport barrier could explain this, possibly one associated with a fast zonal flow~\cite{Bowman:1996,Haynes:2005,Rypina:2007,Read:2004,Read:2007,Dunkerton:2008,Wood:2010,Provenzale:1999}. The anti-correlated torque and flow measurements will be discussed
 more in Sec.~\ref{discuss}.

 \section{\label{fluc}Turbulent Flow Fluctuations}
\par The turbulent fluctuations are significantly different in the two flow states. The measured velocity fluctuations in the high torque
state are larger than those in the low torque state, despite the lower
mean velocity. Fig.~\ref{UPDF}$(a)$ shows the probability density of the
dimensionless velocity conditioned on the low and high torque
states, with Gaussian curves for comparison.  The mean azimuthal velocity
measured in the low torque state is 2.45 times that seen in the high
torque state (in agreement with Fig.~\ref{UtauG}), while the fluctuations in the high torque state
are 1.5 times the low torque state fluctuations.  Based on this
measurement, the high torque state has a considerably higher
turbulence intensity  $\sigma_{u^{\prime}}/\langle
u^{\prime}\rangle$, of 57\%. The low torque state turbulence intensity is 16\%.
\par
The wall shear stress fluctuations are similar in magnitude between the two states, as shown in Fig.~\ref{UPDF}$(b)$.
 The low torque state mean wall shear is about 1.5 times that in the
 high torque state, and the standard deviations in the two
 states agree to within 5\%. The high torque state wall shear
distribution is significantly skewed, with a skewness of about
0.6.
\section{\label{waves}Waves}
\begin{figure}[ht]
 \includegraphics[width=8.6cm]{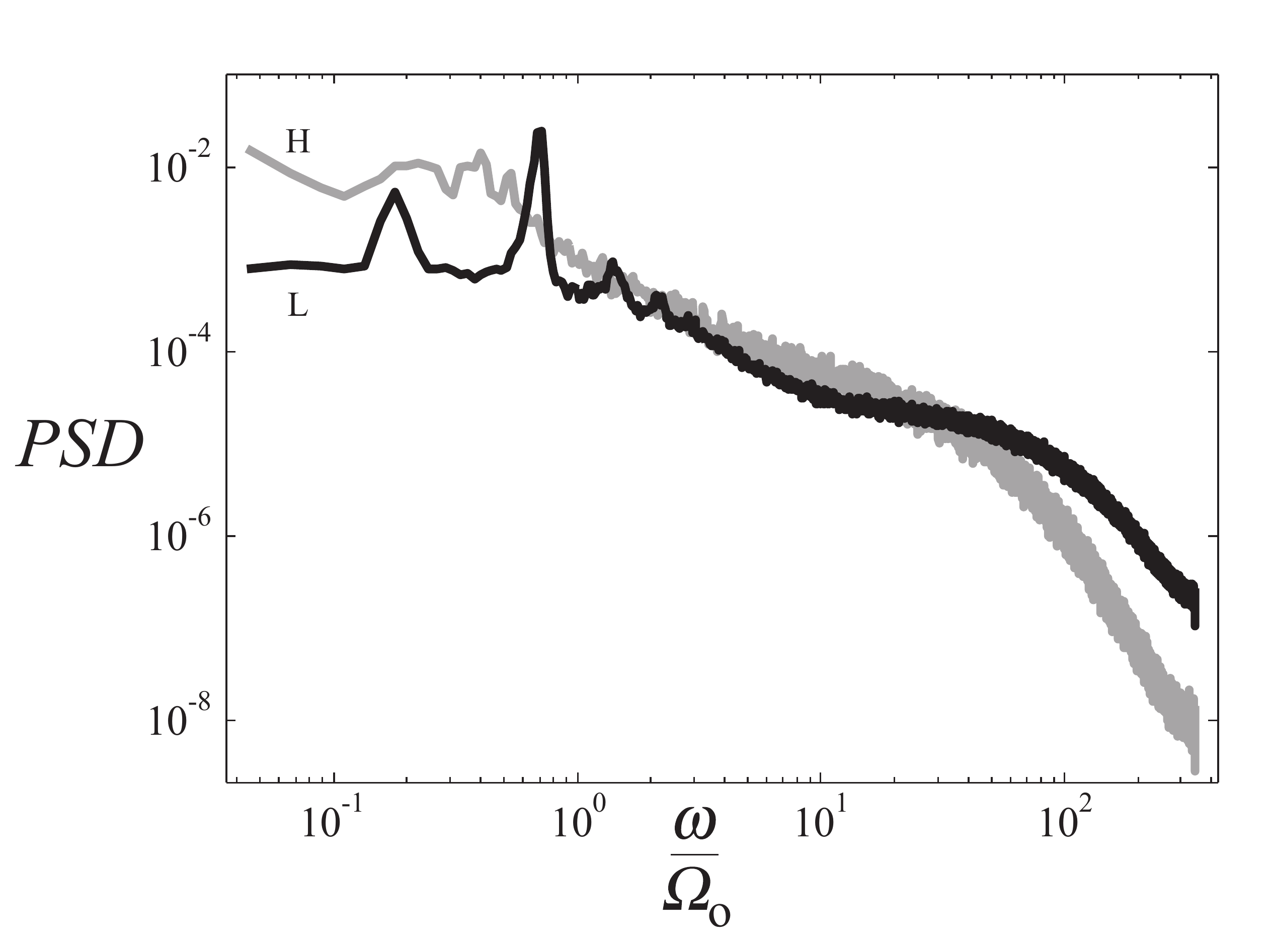}
 \caption{\label{Wspec}{Power spectra of wall shear stress at $Ro = 2.13$ and $E=2.1\times10^{-7}$, conditioned on state.   Angular frequency has been made
 dimensionless using the outer sphere angular speed. The black curve is the spectrum
 from the low torque state, and the gray curve is that of the high torque state.  The low torque spectrum has prominent peaks at $\omega/\Omega_\mathrm{o} = 0.18$, 0.71, and harmonics. In the high torque state there are broad peaks at $\omega/\Omega_\mathrm{o} = 0.40$ and $0.53$.}}
 \end{figure}

Some of the most energetic fluctuations in the low torque state are system-scale
wave motions with well defined frequencies.  The high torque state
has more broadband, low frequency fluctuations. The conditional wall shear
stress power spectra of Fig.~\ref{Wspec} show this clearly.  Two wave motions feature prominently in the low torque state but peaks are nearly absent in the high torque state.  The lower peak in the low state spectrum of Fig.~\ref{Wspec} has a frequency of $0.18\thinspace\Omega_\mathrm{o}$.  Measurements with the multiple pressure sensors show that this wave has azimuthal wavenumber $m=1$.  The higher peak has a frequency of $0.71\thinspace\Omega_\mathrm{o}$ at this Rossby number, and several higher harmonics are also visible.  This higher frequency wave appears to have $m=2$.
\par
\begin{figure}[ht]
 \includegraphics[width=8.6cm]{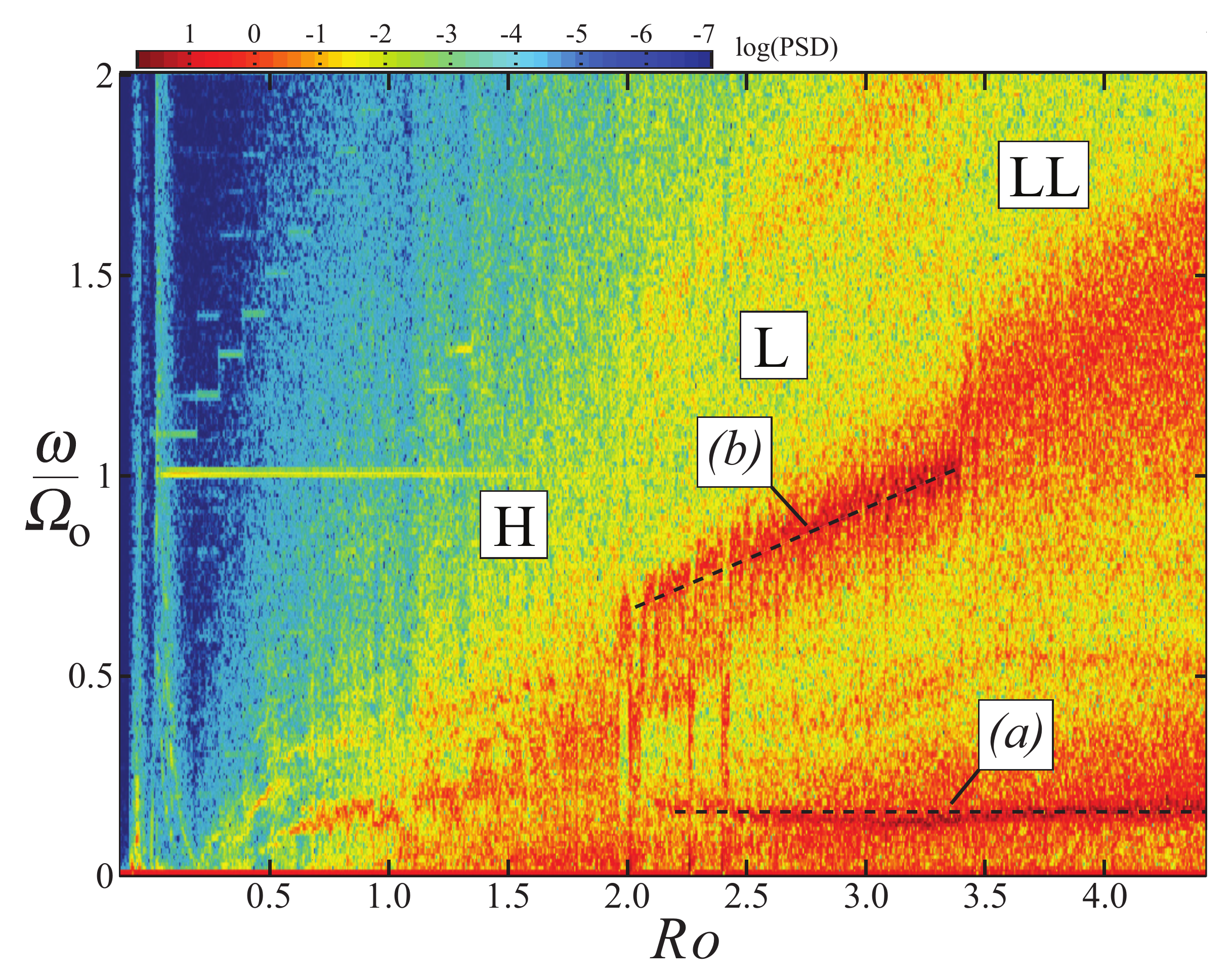}%
 \caption{\label{RPspec}{A spectrogram of wall pressure at $23.5^\circ$ colatitude shows evidence of several flow transitions as $Ro$ is varied.   $E = 1.3\times10^{-7}$ and $0.1\leq$Ro$\leq4.4$, waiting 430 rotations per step with steps of $Ro = 0.1$.  Instead of averaging the power spectra over an entire
 step of $Ro$, there are 30 spectra per step in $Ro$, so some temporal evolution is visible at a given $Ro$.   In the $L$ state, there are two strong waves, the lower, at $(a)$, varies only
 slightly in frequency with $Ro$.  The higher frequency wave at $(b)$ varies more strongly with $Ro$, suggesting that advection by the mean flow is important in setting its frequency. }}
 \end{figure}

The frequency and azimuthal wavenumber of the lower frequency wave are consistent with the full sphere inertial mode~\cite{Zhang:2001} (3,1,-0.1766) in the ($l$,$m$,$\omega/\Omega_\mathrm{o}$) notation of Greenspan~\cite{Greenspan:1968} and Kelley~\emph{et al.}~\cite{Kelley:2007}.    The inertial modes are not known analytically for the spherical shell~\cite{Rieutord:1997,Rieutord:2001,Tilgner:1999}, but there is good
experimental agreement in frequencies and spatial patterns between inertial modes observed in turbulent spherical Couette flow~\cite{Kelley:2007} for $Ro<0$ and modes of the full sphere~\cite{Zhang:2001}. Like the inertial modes observed previously~\cite{Kelley:2007}, the dependence of this mode's frequency on the Rossby number is weak.  The spectrogram of Fig.~\ref{RPspec} shows the variation of pressure power spectra from a single pressure sensor as $Ro$ is varied and $E$ held constant. Line $(a)$ in Fig.~\ref{RPspec} is at $0.16\thinspace\Omega_\mathrm{o}$.  The strong lower frequency peak starts higher, $0.18\thinspace\Omega_\mathrm{o}$, and varies down to $0.14\thinspace\Omega_\mathrm{o}$ and back up to $0.16\thinspace\Omega_\mathrm{o}$ as $Ro$ is increased.  This peak is featured in flow power spectra over a wide range, not disappearing until about $Ro=12$.  It is worth mentioning that  $(3,1,-0.1766)$ inertial mode of the sphere is one of a special class of slow, geostrophic inertial modes that are equivalent to Rossby waves propagating on a solid body background~\cite{Busse:2005}.
 \par
The frequency of the stronger, higher frequency $m=2$ wave varies more strongly with the Rossby number.  The equation for Fig.~\ref{RPspec} line $(b)$ is
\begin{equation}\label{Roline}
\frac{\omega}{\Omega_\mathrm{o}} = 0.25 Ro + 0.16.
\end{equation}
This variation with $Ro$ suggests a Rossby wave that is doppler shifted by advection.  Rossby waves propagate where there is a gradient of potential vorticity, which in an isothermal, incompressible fluid is the quotient of fluid absolute vorticity and fluid column height. In the $\beta$-plane approximation common for Rossby waves~\cite{Rossby:1939,Rossby:1940,Rossby:1945,Dickinson:1978,Barcilon:2001},  the dimensionless frequency of the waves is given by
\begin{equation}\label{Rossfreq}
\frac{\omega}{\Omega_\mathrm{o}} = k \frac{U}{\Omega_\mathrm{o}}-\frac{\beta}{k \Omega_\mathrm{o} },
\end{equation}
where $k$ is the wavenumber and $U$ is the velocity of a mean flow advecting the waves.  The parameter $\beta$ is related to the background potential vorticity gradient on which the waves propagate. In our system, this gradient could result from the topographical effect of the sloping boundaries~\cite{Schaeffer:2005,Schaeffer:2005a} or from a gradient of relative vorticity caused by the mean azimuthal flow profile.
We observe that the zonal flow velocity $U$  varies linearly with the Rossby number $Ro$ in the low torque state.  Outside the tangent cylinder with the inner sphere super-rotating, both the topographical contribution to $\beta$ and the probable contribution from the mean relative vorticity would make it negative, consistent with the positive intercept of Eq.~\ref{Roline}.   Therefore Eq.~\ref{Rossfreq} and Eq.~\ref{Roline} are indeed similar, and suggest an advected Rossby-type wave.
\par
At $E=10^{-3}$ and $Ro=2.9$, in the low torque state range, Guervilly and Cardin observe an $m=2$ Rossby-type wave~\cite{Guervilly:2010}.  They do not report an additional $m=1$ wave. However, if the physical mechanism that gives rise to the $m=2$ mode there is robust to lowering the Ekman number from $10^{-3}$ to $10^{-7}$, it could be that the $m=1$ wave is excited by the $m=2$ advected wave when the inertial mode damping is low enough.  Reynolds stresses from nonlinear waves in rotating systems can transport angular momentum by driving strong zonal flows~\cite{Zhang:2004,Onishchenko:2004,Plaut:2008,Tilgner:2007,Morize:2010}, and the strength of these zonal flows compared to the wave amplitude grows rapidly as the Ekman number is decreased~\cite{Tilgner:2007}.  Therefore, the waves could play an important role in the angular momentum dynamics we observe.

\section{\label{discuss}Discussion}
\begin{figure}[ht]
 \includegraphics[width=8.6cm]{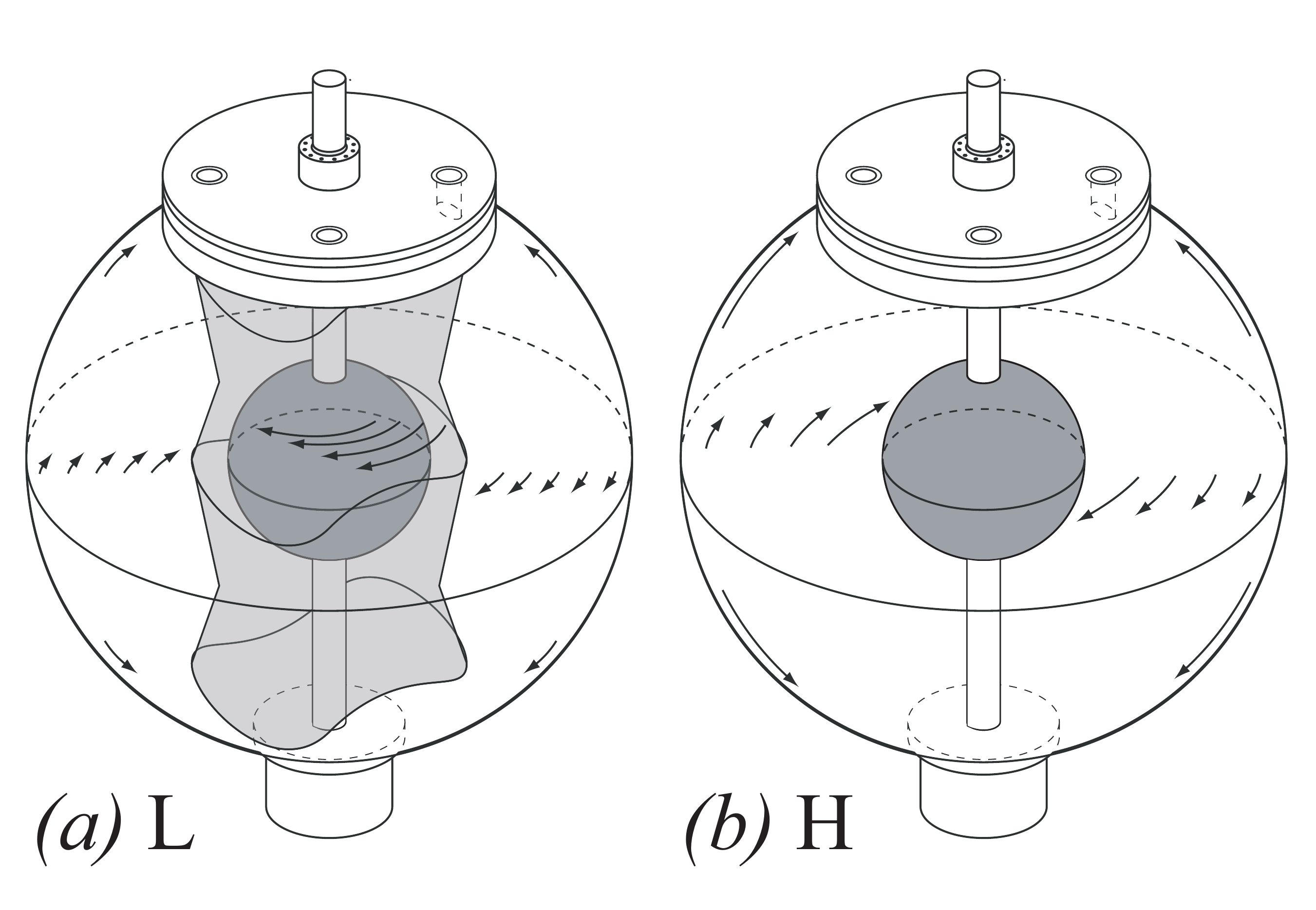}
 \caption{\label{sketch}{A sketch of two possible mean flow states.  The low torque state is labeled $L$ and the high labeled $H$.  The low torque state at $(a)$ is characterized by
 fast zonal circulation near the core of the experiment and large amplitude waves.    In the high torque state, the velocity profile varies more gradually.  The zonal circulation has been destroyed
 by mixing across the transport barrier, so the fluid near the inner sphere is slower and the torque higher.}}
 \end{figure}

One possibility for the state transitions discussed here is the formation and destruction of a fast zonal circulation at small cylindrical radius, with a resilient barrier to transport in the low torque state.  Such barriers are common features in rotating turbulent  flows~\cite{Rypina:2007,Haynes:2005, Bowman:1996,Read:2004,Read:2007,Dunkerton:2008,Wood:2010,Dritschel:2008}.  The
fast mean flow and strong wall shear stress at high latitude in the low torque state, along with the falling total angular momentum of Fig.~\ref{GnL} mean that the two states must have a different shape of the profile of angular momentum with cylindrical radius ($s^2\Omega(s)$).  The  low torque state flow must favor, on average, fast circulation at the center of the experiment and slower at larger radius.  A simple possibility would be a central zonal circulation bounded by a sharp shear layer modulated by the advected waves discussed in Sec.~\ref{waves}.  This is shown schematically in Fig.~\ref{sketch}$(a)$.  A collapse of this shear layer could cause the abrupt low state to high state transition, suddenly surrounding the inner sphere with low angular velocity fluid and increasing the torque.   The very short timescale of the L to H transition, 10 rotations of the outer sphere, suggests that some sort of rapid mixing is responsible.  When the fluid is angular momentum is closer to well mixed, we expect a relaxation to a less steeply varying angular velocity profile, as in Fig.~\ref{sketch}$(b)$.
\par
 In the low torque state schematic, Fig.~\ref{sketch}$(a)$, the azimuthal flow outside the shear layer is slow, but not locked to the outer sphere.  The angular momentum measurements of Fig.~\ref{GnL} suggest that the Ekman suction on the outer boundary drains angular momentum from the region outside the shear layer faster than the flux across the zonal flow boundary can replenish it until an equilibrium is reached or a L to H transition occurs.
\par
At the transition from the high torque to low torque state, the slow decay of the inner torque and increase of the observed fluid velocity and wall shear suggest that the transport barrier bounding the central zonal flow
has re-formed, but the inner sphere must still provide angular momentum and energy to spin up the fast central flow.  At this point, since the exterior fluid is now weakly coupled to the fast inner sphere, it starts to spin down.
\par
We mention finally that changes in mean profiles of angular momentum by vigorous mixing have been studied in rotating thermal convection~\cite{Aurnou:2007,Brun:2009,Gilman:1977}.  Mixing by convectively driven fluctuations tends to flatten the angular momentum profile associated with solid body rotation, and this results in faster velocities in the rotating frame at high latitudes~\cite{Aurnou:2007}.  Increasing vigor of convective mixing of the conserved angular momentum results in a switch from a prograde equatorial zonal jet to a retrograde one in the models of Aurnou~\emph{et al.}~\cite{Aurnou:2007}. Advectively flattened angular momentum in the bulk of turbulent swirling flows is well established in Taylor-Couette flow~\cite{King:1984,Lathrop:1992,Lewis:1999} when the outer vessel is stationary, though the profile is not so flat in outer-stationary spherical Couette~\cite{Sisan:2004a}.
\par
Since the total angular momentum of the system is fluctuating, we must be careful not to use arguments based on strict angular momentum conservation.  Angular momentum is freely exchanged with the rotating walls.  An alternative explanation based on angular momentum mixing from vigorous turbulent fluctuations could involve a profile closer to the quadratic solid body profile in the high torque state and an increasingly flattened angular momentum profile in the low torque state, and would not involve the formation and destruction of a mixing barrier. It does, however, require more than a simple redistribution of angular momentum to explain the results of Fig.~\ref{GnL}.

\section{\label{dynsys}Dynamical Behavior}
\begin{figure}[ht]
 \includegraphics[width=8.0cm]{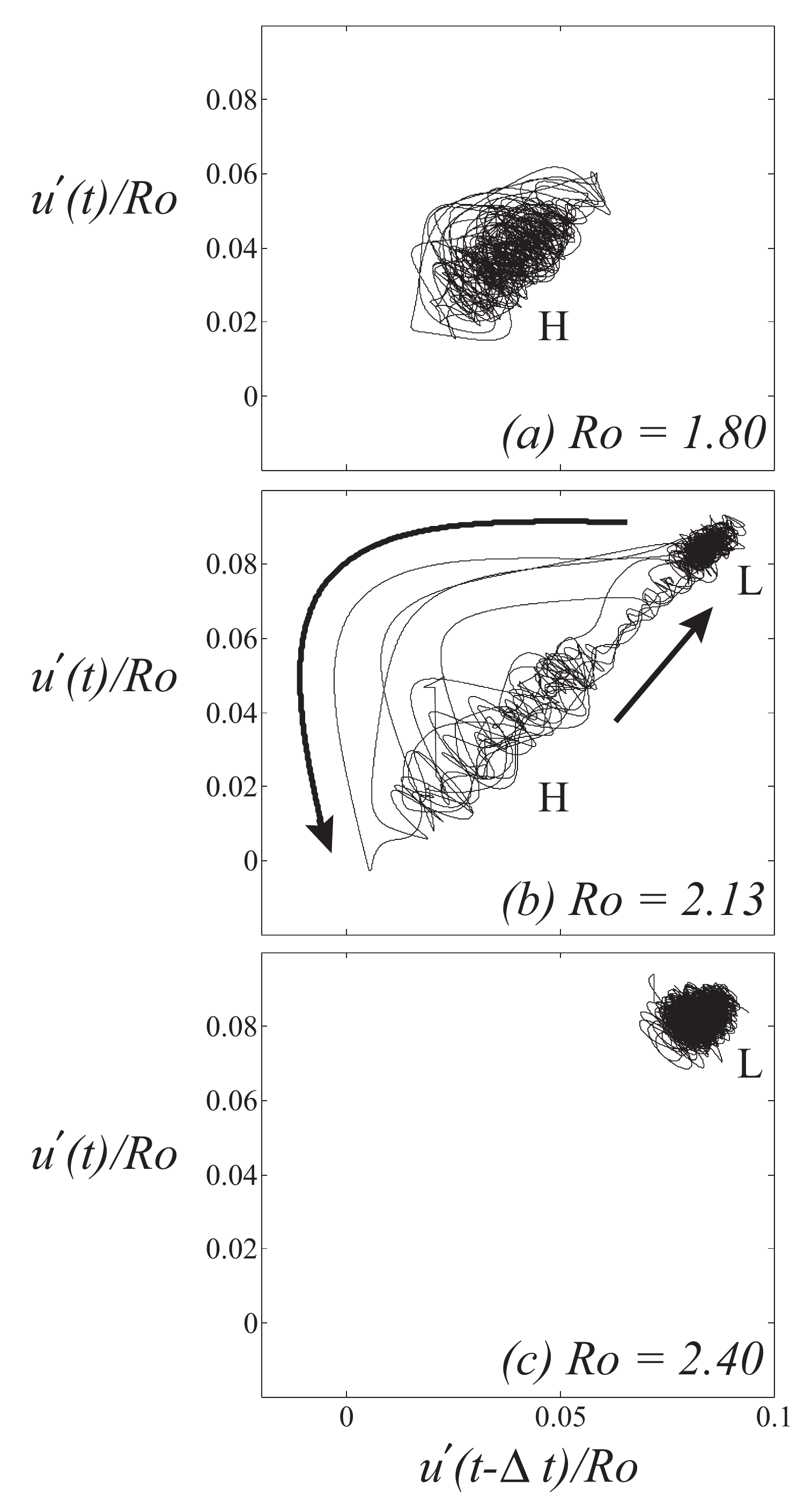}

 \caption{\label{tdel}{Time delay embeddings of the slow velocity fluctuations at $y^{\prime} = 0.02$ for three values of
 $Ro$ below, in, and above the bistable range.  Arrows in $(b)$ show the direction taken by the transitions.
  The time delay $\Delta t$ is 4.5 rotations of the outer
 sphere in all three cases.  The dimensionless velocity as defined
 previously has been scaled by $Ro$, roughly collapsing the mean
 velocity and fluctuation levels of each state.
 These data have been low pass filtered with $f_c =
0.1$~Hz, or a period of 7.5 rotations of the outer sphere.  }}
 \end{figure}
We also investigated the evolution in phase space of H$\leftrightarrow$L transitions using low dimensional time delay embedding of low pass filtered velocity data. Figure~\ref{tdel} is a 2D embedding of a low pass filtered velocity time series.  The filter is a 4th order Butterworth with a cutoff frequency $f_c = 0.1$~Hz. This corresponds to a period of 7.5
rotations of the outer sphere. This velocity signal was plotted
against the same signal 4.5 rotations prior. Figure~\ref{tdel}~$(a)$ is $Ro=1.8$,
the lower threshold of the first bistable range.  At $(b)$, we have $Ro=2.13$, the same as in Fig.~\ref{Gtime}, where the bistable switching is
evident, and in $(c)$, $Ro=2.4$, above the range where spontaneous
transitions are observed.

\par The velocity is rescaled here by the outer sphere tangential
speed to make it dimensionless and by $Ro$, the expected
dimensionless velocity scale relative to the outer sphere.   Figure~\ref{tdel}$(b)$ could be interpreted as a heteroclinic connection between two turbulent
attractors, with connections between the high torque flow state in $(a)$ and the low torque flow state in $(c)$.
The arrows in Fig~\ref{tdel}$(b)$ show the direction of flow in phase space as the
system undergoes several transitions.  Below the critical Rossby number in Eq.~\ref{fitform}, we
expect H$\leftrightarrow$L connection to be broken.  Above the bistable
range, we recall the probability of state as a function of $Ro$ of Eq.~\ref{fitform} and Fig.~\ref{stateprob} and expect that connections between attractors weaken but do not break.

\section{\label{conclusions}Conclusions}
The experimental results presented here reveal novel turbulent multiple stability and turbulent flow transitions in a previously unexplored parameter range of very high Reynolds number,
rapidly rotating spherical Couette flow.  The transitions appear to involve the formation and destruction of zonal flow transport barriers and exhibit
strong waves and unusual angular momentum transport.  These results suggest spherical Couette as another straightforward laboratory testbed for studying turbulent
multiple stability and could help lead to a better understanding of similar phenomena in natural systems.
\par
We are grateful to the NSF Geophysics program for funding and to the two anonymous reviewers for many helpful suggestions.


\bibliography{../../../mainref}
\end{document}